# Phase diagram of Symmetric Iterated Prisoner's Dilemma of Two-Companies with Partial Imitation Rule

**Liangsheng Zhang, Wenjin Chen, Mathis Antony, and K. Y. Szeto***

Department of Physics,

Hong Kong University of Science and Technology,

Clear Water Bay, Hong Kong, HKSAR, *Corresponding author email: phszeto@ust.hk

## Abstract

The problem of two companies of agents with one-step memory playing game is investigated in the context of the Iterated Prisoner's Dilemma under the partial imitation rule, where a player can imitate only those moves that he has observed in his games with his opponent. We limit our study to the special case where the players in the two groups enjoy the same conditions on a fully connected network, so that there are only two payoff matrices required: one for players playing games with members of the same company, and the other one for players playing games with members from a different company. We show that this symmetric case of two companies of players can be reduced to the one-company case with an effective payoff matrix, from which a phase diagram for the players using the two dominant strategies, Pavlov and Grim Trigger can be constructed. The phase diagram is computed by numerical integration of the approximate mean value equations. The results are in good agreement with simulations of the two-company model. The phase diagram leads to an interesting conclusion that a player will more likely become a Grim Trigger, regardless of their affiliated company, when the noise level increases so that he is more irrational, or when the intra-group temptation to defect increases.

PACS number(s): 02.50.Le, 87.23.Kg, 87.23.Ge, 87.19.lv

# I. Introduction

Evolutionary game dynamics has been a topic studied intensively in the past few decades [1]-[7], with many applications in biology, ecology and economics [5][8]-[10]. Among numerous models for the game, the most studied one is the Iterated Prisoner's Dilemma (IPD) [4]-[7], in which players have a chance to play games repeatedly. Players may reward or punish their opponents based on their actions in previous rounds. Given a chance to revise a player's strategy, imitation has been an important means of both strategy learning for individual players and strategist evolution for the whole population [11][12]. For players with one step memory in the IPD game, several important strategies, such as Tit-for-Tat (TFT), Pavlov and Grim Trigger (GT), have been studied extensively and they dominate over other strategies in many scenarios [2][4][12]. However, recent studies have revealed that the traditional imitation process involves unrealistic assumptions on the players' ability to imitate their opponents' moves that are not even been observed. This observation initiates the introduction of the partial imitation process, which is to take into consideration the incomplete information of the players. During the imitation process, a player can only imitate those moves that he has observed, not the whole meta-strategy of his opponent. Such an imitation process is called *partial imitation* [13]-[15]. Simulation based on this partial imitation rule (pIR) has shown very different fraction of strategists in the steady state, as compared to the traditional imitation rule (tIR), where the entire set of moves of the opponent will be copied. The differences between tIR and pIR have been extensively investigated using numerical simulation as well as *approximate mean value equations* [15], which agree well with numerical simulation.

In this paper, we extend the analysis of the single population of players with one-step memory, imitating according to the partial imitation rule, to two competing populations of players. In the context of econophysics, these two populations of players correspond to agents belonging to two different companies. If we label the two populations of players by their companies, A and B, in general, we anticipate that there are three payoff matrices required to describe their evolution. The first one is a matrix $M_{AA}$ that determines the payoff between players in company A. The second matrix $M_{BB}$ determines the payoff between players in company B. A third matrix $M_{AB}=M_{BA}$ determines the payoff between two players coming from different companies. In this paper, we make the simplifying assumption that $M_{AA}=M_{BB}=M_{intra}$, and $M_{AB}=M_{BA}=M_{inter}$, where the subscripts *intra* and *inter* stand for "same company" and "different company", respectively. We call this special case the symmetric case of the two-company model, which characterizes the situation where people sharing a common identity or culture treat strangers differently from their partners. From an earlier paper that addresses this scenario using the Ising model [16], we have found that there exists a way to reduce the analysis of the two-company model using a one-company model with an effective interaction parameter. We find that this paradigm does carry over from the Ising model to the 2x2 game of two companies of players.

Based on the partial imitation process, this paper addresses the possible phase transition between the Grim Trigger dominant regime and the Pavlov dominant regime. Using the approximate mean value equations for the imitation process of a single company of players with an effective payoff matrix, we find good agreement between numerical simulations of the original two-company model with the integration of the mean value equations for the one-company model. The analysis reveals the existence of a phase transition between the two dominant types of players, Grim Trigger and Pavlov, with a three-dimensional phase diagram parameterized by the noise factor in the imitation process, the "inter-" and the "intra-" company payoff matrix.

**II. Model**

**IIA. Payoff Matrix**

In a Prisoner's Dilemma game between two players, each person chooses either to cooperate (*C*) or to defect (*D*). Based on these two choices the two players receive their payoff. A cooperating player scores *R* (*S*) if his opponent cooperates (defects) and a defecting player scores *T* (*P*) if his opponent cooperates (defects). Here *R* stands for the Reward for cooperation, *S* the Sucker's payoff, *T* the Temptation to defect and *P* the Punishment for mutual defection. The Prisoner's Dilemma game imposes the following restrictions on the payoff parameters: $T > R > P > S$, and to prevent collusion $2R$ should be greater than $T + P$. The game between players from two companies A and B in general have three possible different payoff matrices: two for intra-company interaction, $M_{AA}$ and $M_{BB}$ and one for inter-company interaction, $M_{AB}=M_{BA}$. All three payoff matrices will adopt the following form

$$\begin{pmatrix} R & S \\ T & P \end{pmatrix} = \begin{pmatrix} 1 & 0 \\ b & 0 \end{pmatrix} \qquad (1)$$

where $b$ (>1) is the temptation to defect and is the only tunable variable, and we assume the case of weak Prisoner's dilemma game [5][17]. For clarity, $b_{intra,A}$ and $b_{intra,B}$ will be used to denote the temptations in the intra-company payoff matrices of population A and B, and $b_{inter}$ will be used to denote temptation in the inter-company payoff matrix. Generally, the three parameters, $b_{intra,A}$, $b_{intra,B}$ and $b_{inter}$ are different for the three payoff matrices. As we limit our study to the special case where the players in the two companies have the same payoff matrix, we have $b_{intra,A} = b_{intra,B} = b_{intra}$. Besides, $b_{inter}$ is required to be no less than $b_{intra}$ in order to characterize the aforementioned hostility-against-stranger behavior.

**IIB. One step memory**

For IPD with one step memory, each strategy can be specified by the response (*C* or *D*) a player will take according to the last round played. Furthermore, the game requires the specification of the initial move of the players. In each round of game between two players, there are four possible outcomes

(("my move"  "opponent's move")=*DD, DC, CD, CC*) with the immediate payoffs *P, T, S* and *R*, respectively. In the context of one step memory, a player can recall his opponent's and his own strategy in the past one round. They can have responses in terms of strategy $S_P$, $S_T$, $S_S$ and $S_R$ for the *DD, DC, CD* and *CC* in the previous step respectively. Together with the initial move $S_0$, we can encode the strategy of IPD with one step memory by the following notation: $S_0|S_P S_T S_S S_R$. For example, GT is $C|DDDC$ and Pavlov is $C|CDDC$. Since a strategy is denoted by $S_0|S_P S_T S_S S_R$ which have five bits and each slot encoded by either *C* or *D*, therefore, there are a total of 32 possible strategies in the strategy space $M_s$ of IPD with one step memory.

**IIC. partial Imitation Rule (pIR) and recurrent state**

The essential idea of pIR is that a player can only imitate the exposed part of his opponent's meta-strategy. For example, let's consider Alice using strategy $D|DDDD = S_0^A|S_P^A S_T^A S_S^A S_R^A$ to play against Bob who uses $C|DCDC = S_0^B|S_P^B S_T^B S_S^B S_R^B$. Alice and Bob will play according to their and their opponents' action in the previous step. Initially, Alice will play *D*, and Bob will play *C*. Thus, for the next step, Alice will see that the pattern for ("my move" "opponent's move") in the previous round is (*DC*), thus her response for the present step should be $S_T^A =D$. For Bob, he will see that the pattern for ("my move" "opponent's move") in the previous round is (*CD*), therefore, his response for the present step should be $S_S^B = D$. Now, in the next step, both Alice and Bob will see the strategy pattern in the previous step to be (*DD*). Thus, they will be both using *D* in response and that will persist forever. To summarize, Alice observes the following transition: after the initial outcome $DC = S_0^A S_0^B$, the second outcome is $DD = S_T^A S_S^B$, and then all subsequent outcomes will be $DD = S_P^A S_P^B$. Therefore, Alice has *only* witnessed Bob using three moves, $S_0^B$, $S_S^B$, and $S_P^B$, which are all the moves she will adopt if she is to imitate Bob. Consequently, Alice's strategy will transform from $S_0^A|S_P^A S_T^A S_S^A S_R^A = D|DDDD$ to $S_0^B|S_P^B S_T^A S_S^B S_R^A = C|DDDD$ if Alice imitates Bob, but this strategy is not exactly Bob's strategy. In fact, Alice will not learn Bob's cooperative moves, which are $S_T^B = C$, and $S_R^B = C$, since Bob has never used these moves in his encounter with Alice. Therefore, Alice retains her own meta-strategy $S_T^A = D$ and $S_R^A = D$ This example shows the difference between the idea behind our partial imitation rule and the traditional imitation rule where Alice's strategy will change from *D|DDDD* to $C|DCDC$.

In the limit of weak selection [8][18], players play a larger number of games before imitation happens. Eventually, players get into the so-called recurrent state [12][15] in which a set of strategy pattern for ("my move" "opponent's move") are revisited over and over again. In the above example, the recurrent state for Alice and Bob is (*DD*). We assume that in an encounter, the game plays long enough so that the initial transient state is negligible. Under this assumption, the average payoff $U_{ij}$ in the encounter between players *i* and *j* is the total recurrent state payoff divided by the number of rounds in the recurrent state. For instance, the payoff for Alice and Bob in the above example is the payoff of the recurrent state (*DD*). If we use the payoff matrix in Eq.(1), both Alice and Bob get payoff zero in the recurrent state.

After obtaining the recurrent state payoff $U_{ij}$ in the encounter between players *i* and *j*, we now compute the total average payoff of a given player. We note that this chosen player will play games with all other players and his total average payoff $U_i$ is

$$U_i = \sum_{j=1}^{N_s} \left( \rho_{j,intra} U_{ij}^{intra} + \rho_{j,inter} U_{ij}^{inter} \right) \quad (2)$$

where $U_{ij}^{intra}$ is the average payoff that a player with strategy *i* obtains when he confronts another player with strategy *j* from the same company, while $U_{ij}^{inter}$ is the average payoff when the other player comes from a different company, $\rho_{j,intra}$ is the density of players with the *j*-th strategy in the same company as player *i*, $\rho_{j,inter}$ is the density of players with the *j*-th strategy in a different company from player *i*'s company, and $N_s = 32$ is the number of strategies in the one-step memory strategy space. Here *j* runs from 1 to $N_s$ so that all strategies are included. Based on the average total payoff, we can proceed to implement the partial imitation of strategies. In general, a player affiliated with a particular company can also switch to the other company, but here we forbid such change for simplicity. We find that in the symmetric two-company model, this extra restriction on the players in their switching of affiliation hardly makes any difference in the final analysis.

**IID. Macroscopic Dynamics**

Realistic players are not totally rational [19]-[21]. The players' irrationality can be characterized by a probability of imitation that depends on the difference of the average payoff calculated in Eq.(2) between the imitator *i* and the role model *j*. Mathematically, if we denote $\Delta U = U_j - U_i$ to be the payoff difference between player *i* and *j*, then the probability of *i* imitating *j* is given by

$$P(i \text{ imitates } j) = g(\Delta U) = \frac{1}{1+\exp\left(-\frac{\Delta U}{K}\right)} = \frac{1}{1+\exp\left(\frac{U_i - U_j}{K}\right)} \quad (3)$$

where *K* is a temperature-like parameter called *noise factor* that controls the degree of irrationality [19]-[21]. Note that the payoff for player *i* and *j* depends on the density of the strategy that player *i* and *j* use. However, the evolution of density of the strategies also depends on the imitation probability in the evolution of the populations. One approach to solve this problem is to use the self-consistent *approximate mean value equations*, which describe the macroscopic dynamics of the population evolution [11][19]-[21]. Based on the traditional imitation rule, the approximate mean value equations

are:

$$\frac{d\rho_i^{tIR}}{dt} = \rho_i \sum_j \rho_j \left[ g(U_i - U_j) - g(U_j - U_i) \right] \quad (4)$$

$$i = 1, ..., 32$$

We recently extend the approximate mean value equations to the case under partial imitation rule [13-15]:

$$\frac{d\rho_i^{pIR}}{dt} = \sum_j \rho_j \sum_k \rho_k g(U_j - U_k) p(k, j, i) - \rho_i \sum_j \rho_j g(U_j - U_i) \quad (5)$$

$$i = 1, ..., 32$$

where $p: M_s \times M_s \times M_s \to [0,1]$ is the probability $p(k,j,i)$ that $k$-strategist will become an $i$-strategist by imitating a $j$-strategist and $M_s$ is the strategy space of one-step memory. In the case of pIR,

$$p(k, j, i) = \begin{cases} 1 & \text{if } k\text{-strategist imitating} \\ & j\text{-strategist becomes} \\ & i\text{-strategist} \\ 0 & \text{otherwise} \end{cases} \quad (6)$$

Given the initial distribution of the strategist density $\rho_i(t=0; i=1,...,32)$, the integration of the self-consistent approximate mean value equations describes the evolution of the strategists. In this paper, we focus on the behavior of the population at steady state. Since GT and Pavlov [13] are the most interesting and dominating strategies in a large range of temptation values $b$ and noise factor $K$, we will only discuss the phase transition associated with these two strategies in this paper.

**IIE. Numerical simulation**

The number of players in the simulation is 500,000. Initially, these players are divided equally to each of the two companies. In each company, all the 32 strategies are present in the same fraction. In the evolution, we define one Monte Carlo (MC) step to have 500,000 imitation processes. As we have assumed fully connectedness among all the players, in each encounter two players are selected randomly either from the same company or from different companies during the imitation process. The imitation probability is determined by Eq.(3). People cannot change from the present company to the other company. The simulation lasts until all of the strategies in both companies reach steady state.

## III. REDUCTION TO ONE-COMPANY CASE

The symmetry introduced in the two-company model refers to the assumption that $b_{intra1} = b_{intra2}$. We show in this section that this provides a simplification that reduces the two-company case to a one-company problem. First, we expect that this symmetry suggests that company imitation is not important since the proportion of players in each company will deviate only a little bit from one half. Let's assume that every strategy in each company will have almost the same fraction, that is $\rho_{iA} \approx \rho_{iB}$ for a certain strategy $i$, where $\rho_{iA}$, $\rho_{iB}$ means the fraction of the strategy $i$ in company A and B respectively, and the symbol $\approx$ indicates that the symmetry is only approximate in the stochastic process of imitation. This assumption will be tested by simulation in Section IV. Nevertheless, if the deviation from the perfect symmetry can be neglected, we assume that the company affiliation does not affect the total average payoff of a certain strategy. Therefore, under perfect symmetry, $\rho_i / 2 = \rho_{iA} = \rho_{iB}$ we can write Eq.(2) as

$$U_i = \sum_j \frac{\rho_j}{2} \left( U_{ij}^{intra} + U_{ij}^{inter} \right) \quad (7)$$

It can be easily verified that if $\vec{C}(Cooperate) = \begin{pmatrix} 1 \\ 0 \end{pmatrix}$ and $\vec{D}(Defect) = \begin{pmatrix} 0 \\ 1 \end{pmatrix}$, then the payoff of player $i$ with meta-strategy $S_i \in \{\vec{C}, \vec{D}\}$ confronting play $j$ with meta-strategy $S_j \in \{\vec{C}, \vec{D}\}$ will be $S_i^T M S_j$ where $T$ is the transposition on vector, and $M$ is the intra or inter-company payoff matrix in Eq.(1). Borrowing Dirac's notation in quantum mechanics, the payoff can be written in a compact way as $\langle S_i | M | S_j \rangle$. The average payoff of player $i$ confronting player $j$ will be

$$U_{ij} = \lim_{Time \to \infty} \frac{1}{Time} \sum_{t=1}^{Time} \langle S_i(t) | M | S_j(t) \rangle = \overline{\langle S_i(t) | M | S_j(t) \rangle} \quad (8)$$

where $t$ is the MC step in the imitation process and $S_i(t), S_j(t) \in \{\vec{C}, \vec{D}\}$ depend on the given strategies $i$ and $j$. Using $M_{intra}$ to denote the intra-company payoff matrix and $M_{inter}$ for inter-company payoff matrix, we can combine Eq.(7) and (8) to write

$$U_i = \sum_j \left[ \frac{\rho_j}{2} \left( \overline{\langle S_i(t) | M_{inter} | S_j(t) \rangle} + \overline{\langle S_i(t) | M_{intra} | S_j(t) \rangle} \right) \right]$$

(9)

Since $S_i(t)$ and $S_j(t)$ are independent of the payoff matrix, Eq.(9) can be simplified as

$$U_i = \sum_j \frac{\rho_j}{2} \overline{\langle S_i(t) | M_{inter} + M_{intra} | S_j(t) \rangle}$$
$$= \sum_j \rho_j \overline{\langle S_i(t) | M_{eff} | S_j(t) \rangle} \tag{10}$$

where

$$M_{eff} = (M_{inter} + M_{intra})/2 \tag{11}$$

is the effective payoff matrix, including the contribution from the intra-company and inter-company payoff matrix, which can be explicitly written in matrix form as:

$$\begin{pmatrix} 1 & 0 \\ b_{eff} & 0 \end{pmatrix} = \begin{pmatrix} 1 & 0 \\ (b_{intra} + b_{inter})/2 & 0 \end{pmatrix} \tag{12}$$

We see that Eq. (12) corresponds exactly to the payoff matrix for a one-company model with temptation $b_{eff}$. Now, according to our assumption that symmetry is preserved on the average in the two-company model, the only difference between the dynamics of a two-company model with the one company model is the way the average payoff is calculated. The calculation involves two different payoff matrices in the former case while in the latter case only involves one. Therefore, we can use the one-company model with effective payoff matrix in Eq. (12) to approximate the symmetric two-company model with reduced computational effort. We will evaluate this approximation using simulation in the next section. Note that this reduction can be generalized to multi-company case as long as the symmetry approximation remains valid. Such reduction to one-company model allows us to use the approximate mean value equation Eq.(5)(6) for the analysis of the multi-company model, which saves computational effort greatly.

## IV. SIMULATION RESULTS AND PHASE DIAGRAM

First, let's examine the low noise regime, with $K$=0.01, so that the imitation process is more rational and the results converge faster towards the steady state. We fix the value of $b_{intra}$ at 1.1 and vary $b_{inter}$ and compare the direct simulation results for the two-company model and the simulation results derived from the effective one-company model. Since the approximate mean value equations have been verified to give good agreement to the simulation result for the one-company model [11][19]-[21], we can therefore directly compare the simulation results of the symmetric two-company model with the integration results from the mean value equations for the one-company case with payoff matrix in Eq.(12). This comparison is shown in Fig. 1. Since the model assumes symmetry on the two companies, we only show the total fraction of a particular strategy $j$, as discussed in section III. The circles and triangles are the simulated steady state strategy density for GT and Pavlov in the symmetric two-company model. The solid and dashed lines are the integrated steady state strategy density for GT and Pavlov using the mean value equations for the one-company model with effective payoff matrix in Eq.(12).

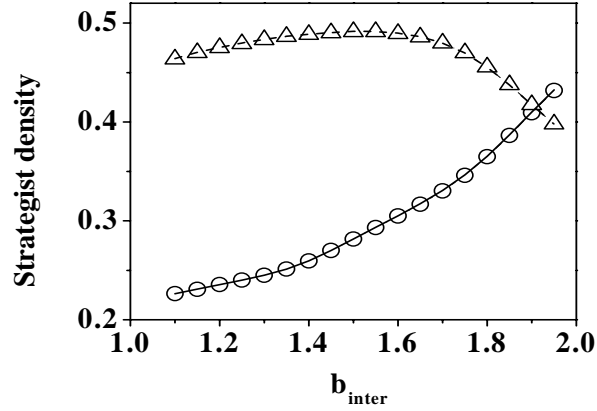

Fig. 1 Total fraction of strategy GT and Pavlov versus $b_{inter}$ when $b_{intra}$ = 1.1 and the noise factor $K$ = 0.01. The circles (triangles) are the strategy densities for GT (Pavlov) in the steady state from numerical simulation, while the solid (dashed) lines are the strategy densities for GT (Pavlov) in the steady state from integration using Eq.(5) with effective payoff matrix in Eq.(12).

We see that the agreement is quite good. We also observe that when $b_{inter}$ increases, Pavlov loses its dominance to Grim Trigger. Now, let's discuss the critical value of $b_{inter}$ at the crossover point of these two strategies and denote it by $b^c_{inter}$. This crossover point marks the outgrowth of these two opposite trends. When $b_{inter} < b^c_{inter}$, we have a Pavlov-dominant phase, and $b_{inter} > b^c_{inter}$ the GT-dominant phase.

By varying $K$ and $b_{intra}$, we obtain a three-dimensional phase diagram, showing the critical value $b^c_{inter}$ as a function of $K$ and $b_{intra}$ subjected to the constraint $1 \le b_{intra}, b_{inter} \le 2$. In order to avoid computational effort in simulation at various set of parameters, we obtain the phase diagram by integrating the approximate mean value equations (Eq.(5)(6)). We first show in the left axis of Fig. 2, the good agreement between the results from integration based on Eq.(5)(6) and the results from numerical simulation on the two-company model at a fixed $K$. The open squares represent $b^c_{inter}$ from the numerical simulation, and the solid line represent $b^c_{inter}$ from the relation $b^c_{inter} = 2b'_{eff} - b_{intra}$, where $b'_{eff}$ is the critical value of effective temptation in the one-company model, obtained through the integration of the mean value equations. Note that $b'_{eff} = b'_{eff}(K)$ is a function of noise $K$, as described by Fig. 2(b). The right axis of Fig. 2(a) shows percentage error $e$ between the numerical simulation of the two-company model and the integration result of the mean value equations: $e = \left(b^c_{inter}(\text{simulation}) - b^c_{inter}(\text{integration})\right)/b_{intra} \times 100\%$. The error is in general below 0.5%. The agreement between the critical value $b_{inter}^c$ for simulation and integration confirms numerically that the simplification of the two-company model to the one-company model in the case of identical intra-company payoff matrix, provides a good approximation, as discussed in Section III.

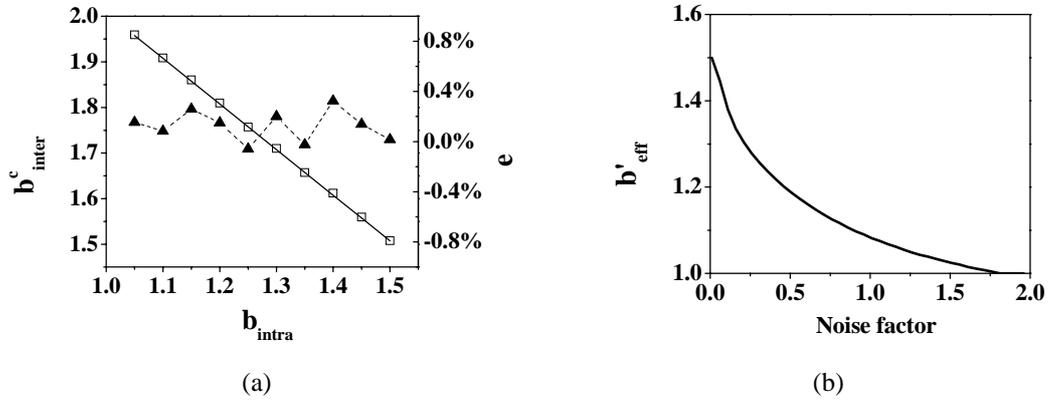

(a)            (b)

Fig. 2 (a) The left Y axis shows the comparison between the results based on numerical simulation on the two-company model (open squares), and the result obtained from the integration of the approximate mean value equations (solid lines). The right Y axis shows in triangles the percentage error *e* of $b_{inter}^c$ between the numerical simulation of the two-company model and the integration result of the mean value equations. (b) Relation between $b_{eff}'$ and noise K obtained through the integration of the approximate mean value equations.

Based on the excellent agreement between the simulation results on two-company model and the integrated results from the approximate mean value equations for the one-company model with effective payoff matrix, we obtain the full phase diagram shown in Fig.3.

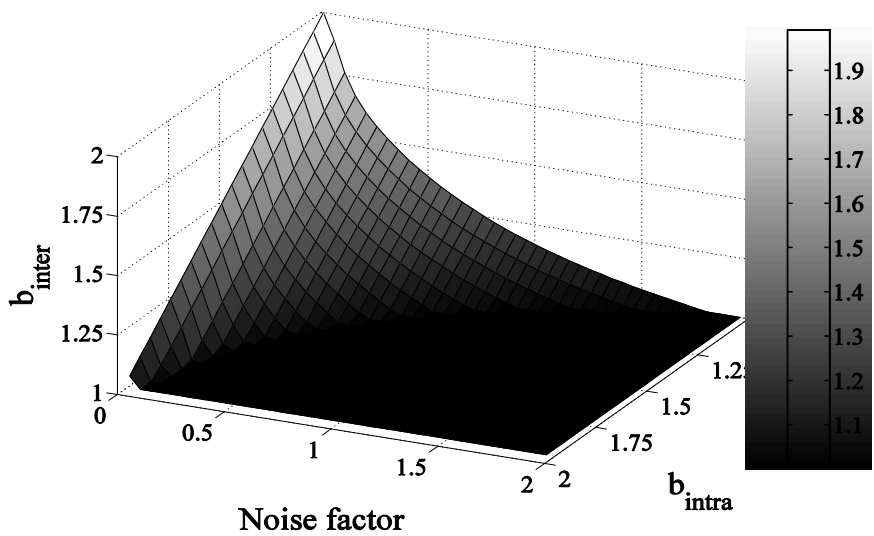

Fig. 3 Phase diagram showing the surface that separates the GT dominant phase (above the surface) from the Pavlov dominant phase.

We notice in Fig. 2 that the linear relation between $b_{intra}$ and $b_{inter}$ at a fixed $K$ in the phase diagram is exactly the one shown in Eq.(11), since the reduction from two-company to one-company is valid. The line defined by

$$b^c_{inter}(K, b_{intra}) = 2b'_{eff} - b_{intra} \qquad (13)$$

can be interpreted as the phase boundary in the following discussion.

From the phase diagram in **Fig. 3**, we see how the domination of GT and Pavlov changes with the parameter $b_{intra}$ and the noise factor $K$. At a fixed $b_{intra}$ and $K$, the GT-dominant region lies in the region of higher $b_{inter} > b^c_{inter}$. Since we have already shown that the reduction to one-company model is good, we can get more insight using the one-company model, which we have a better understanding. In general, a high $b$ value in a one-company model corresponds to a more exploiting game, i.e., locally it favors defection rather than cooperation, as the temptation to defect is high. Meanwhile, GT can be considered as a more defensive strategy than Pavlov (GT player only cooperates when both of the players cooperates in the previous round of game while Pavlov player cooperates when both of them defects or cooperates in the previous round of game), although both strategies are considered cooperative (They both do not seek to defect at the first place; when they and their opponents both play C in the last move they will seek to continue to play C in the current move). In the context of the two-company model, when there is sufficient animosity between two companies, i.e., in the high $b_{inter}$ region, all players will tend to defend their payoff by playing GT, as shown in Fig. 1. We see that the critical value for GT to be dominant is $b^c_{inter}$, which is decreasing with increasing $b_{intra}$ at a given fixed noise level $K$, which in turn also fixes $b'_{eff} = b'_{eff}(K)$. Thus, in the presence of fierce competition or animosity for players of the same company (large $b_{intra}$), the critical $b^c_{inter}$ will be reduced, making the player more inclined to become a Grim Trigger player. In another word, for a given $b_{inter}$ that is initially below the original critical value $b^c_{inter}(K, b_{intra})$, an increased value of $b_{intra}$ that leads to a new smaller value $b^{c,NEW}_{inter}$ can result in having $b_{inter}$ larger than $b^{c,NEW}_{inter}$. Thus, increasing $b_{intra}$ sufficiently can lead to the dominance of Grim Trigger over Pavlov. This leads to the following interesting interpretation. The more inimical players are towards others in the same company, the less animosity from another group they can tolerate. In this case, the player becomes even more hostile and is more likely to defect when he encounters a player of the opposing company, when hostility with players of his own company increases.

We can also analyze the critical $b^c_{inter}(K, b_{intra})$ as a function of the noise $K$. From Fig.2b, the larger the noise, the smaller the $b'_{eff}$. This means that for fixed $b_{intra}$, the increased noise level will reduce the critical $b^c_{inter}(K,b_{intra})$. Again, for a given $b_{inter}$, which may originally be less than a given $b^c_{inter}(K,b_{intra})$ and Pavlov is dominant, an increased noise reduces $b^c_{inter}$ to $b^{c,NEW}_{inter}$ so that $b_{inter}$ can become larger than $b^{c,NEW}_{inter}$ and the dominant strategies can be changed from Pavlov to GT.

Thus, noise or irrationality, tends to agitate players to become more hostile and defensive to others indiscriminately, e.g. people in a circle with not quite rational friends (their friends changes their normal behavior occasionally) tends to be more defensive. The result again is that they tend to become GT and are more likely to defect.

## IV. CONCLUSION

In this paper, we modify the traditional Prisoner's Dilemma game to become more reasonable modification in two ways. Firstly, we adopt the partial imitation rule between players with one-step memory, so that a player only imitates those moves observed during games with his opponents, not those that have not been observed. This realistic modification has been discussed in several recent papers [13-15]. The second modification of the Prisoner's Dilemma game comes from the natural clustering of players to form groups, so that they incline to be more cooperative (less hostile) to their colleagues than to a player of the opposite camp. This is a natural extension of a homogeneous group of players, and can be interpreted in the context of econophysics for agents belonging to two competing companies. This introduction of companies or groups takes into account some inherent difference among players during the payoff calculation. With these two modifications, we address the simple case of only two groups of players who have different level of temptation to defect, depending on affiliation of his opponent. We find that when the two groups of players have similar defensive level among their colleagues, in the sense that they have the same intra-group temptation to defect, the two-company Prisoner's Dilemma game can be mapped onto a one-company model with an effective payoff matrix, characterized by an effective temptation to defect. The mapping is approximate, but numerical simulation of the game shows that this approximation is good. Furthermore, using the mean value equations for the evolution of the strategy density, we find that there is a crossover of dominant strategies between the Grim Trigger and Pavlov. The mean value equations also enable us to draw a phase diagram, with a surface in the three-dimensional parameter space of noise, inter-company and intra-company temptation. The interpretation of this phase boundary, defined by $b^c_{inter}(K,b_{intra})$ in Eq.(12), reveals the interesting result that a player will become more hostile and defensive to others, regardless of their affiliated company when the noise level increases, i.e. when he is more irrational. This also happens when the intra-group temptation to defect increases. In summary, Grim Trigger will more readily dominate Pavlov with either increased noise or increased intra-company temptation to defect.

Finally, it is important to remember that all the results in this paper are based on the assumptions that the players have one-step memory, learning from opponents using the partial imitation rule, and that the two companies of players have identical intra-company temptation to defect. Our future work will investigate the validity of our conclusions when any or all of these assumptions fails. For example, we can ask if the results remain valid when the players have longer memory. We anticipate that this extension is a computational nightmare, since the dimension of the space of strategies increases exponentially with the length of memory. Recent studies on Prisoner's Dilemma game with two-step memories have already shown to be very difficult [12]. Further extension to longer memory could be even more difficult, as the number of possible strategies amounts to $10^{21}$ in the three-step memory case [13]. On the other hand, it is more interesting, hopefully also easier, to investigate the question of symmetry breaking process on the intra-company temptation parameter, $b_{intra}$. New phenomena may emerge when $b_{intra,A}$ for company A is sufficiently different from $b_{intra,B}$ for company B. The phase diagrams for these cases are certainly much more complex.

## Acknowledgements

K.Y. Szeto acknowledges the support of grant CERG602507.